\documentclass[12pt]{article}

\newcommand{\bbr}{I\!\! R}

\newcommand{\x}{arXiv:}
\newcommand{\m}{\mathrm}
\newcommand{\be}{\begin{equation}}
\newcommand{\ee}{\end{equation}}
\newcommand{\ba}{\begin{eqnarray}}
\newcommand{\ea}{\end{eqnarray}}

\usepackage{graphicx}
\usepackage{amssymb}
\usepackage{amsmath}
\usepackage[T1]{fontenc} 
\usepackage[ansinew]{inputenc} 
\usepackage[nosort]{cite}
\newcommand{\inbar}{\vrule height1.57ex width.4pt depth0pt}
\newcommand{\SW}{\relax{\hbox{$\ \inbar\kern-.285em{\rm S}$}}}

\begin{document}
\thispagestyle{empty}
\begin{center}

\null \vskip-1truecm \vskip2truecm

{\Large{\bf \textsf{Applied Holography of the AdS$_5$-Kerr Spacetime}}}

{\Large{\bf \textsf{}}}

{\large{\bf \textsf{}}}

\vskip1truecm

{\large \textsf{Brett McInnes
}}

\vskip0.1truecm

\textsf{\\ National
  University of Singapore}
  \vskip1.2truecm
\textsf{email: matmcinn@nus.edu.sg}\\

\end{center}
\vskip1truecm \centerline{\textsf{ABSTRACT}} \baselineskip=15pt
\medskip

Asymptotically Anti-de Sitter Kerr black holes (we focus here on the five-dimensional case) are associated holographically with matter at conformal infinity which has a non-zero angular momentum density. It is natural to attempt to associate this angular momentum with the recently discovered vorticity of the plasmas produced in peripheral heavy-ion collisions. We assume that an AdS$_5$-Kerr black hole with angular momentum to mass ratio $\mathcal{A}$ is dual to boundary matter with an angular momentum density to energy density ratio also equal to $\mathcal{A}$. With this assumption, we find that, for collisions corresponding to a given value of $\mathcal{A}$, there is a maximal possible angular velocity (well below the maximal value permitted by causality) for such matter at infinity, and that this value is in approximate agreement with the experimentally reported value of the average plasma vorticity produced in typical peripheral collisions of heavy ions.

\newpage

\addtocounter{section}{1}
\section* {\large{\textsf{1. Black Hole Angular Momentum and its Holographic Dual}}}
There is a clear sense in which a \emph{generic} electrically neutral astrophysical black hole is represented by the asymptotically flat Kerr metric in four dimensions: the Schwarzschild metric only occurs as an extremely special case. In the five-dimensional asymptotically AdS case, the role of the generic black hole metric (with a topologically spherical event horizon) is played by the AdS$_5$-Kerr metric \cite{kn:hawk}; the corresponding spacetime merits close attention on those grounds alone, and in fact this statement holds true in a more specific sense, as follows.

In its most familiar and best-studied form, the holographic or \emph{gauge-gravity duality} posits that physics in an asymptotically AdS$_5$ spacetime is dual to that of an $\mathcal{N} = 4$ super-Yang-Mills theory, with a large number of colours, defined on the four-dimensional conformal boundary. There is some reason to hope that this kind of field theory can shed some light on the behaviour of the Quark-Gluon Plasmas (henceforth, QGP) produced in collisions of heavy ions \cite{kn:nat}. Since such plasmas equilibrate very quickly and so have well-defined temperatures, one focuses on bulk systems with similarly well-defined temperatures, that is, on asymptotically AdS$_5$ black holes with large Hawking temperatures.

Most collisions of this kind will be measurably peripheral, that is, off-centre to some non-trivial degree (measured by the parameter known as \emph{centrality}.) In such a collision, a large quantity of angular momentum is  transferred to the QGP \cite{kn:liang,kn:bec,kn:huang}, and so the gauge-gravity dual must likewise have a large angular momentum. \emph{Clearly, then, we need to understand the holography of the AdS$_5$-Kerr black holes}: these represent the \emph{generic} case in this application, just as the asymptotically flat Kerr black hole is generic in the astrophysical application\footnote{The gauge-gravity duality for rotating (topologically spherical) AdS black holes has been studied, and successfully applied, previously: see \cite{kn:sonner,kn:schalm}. These works explain in detail how the AdS/CFT correspondence works in the rotating case. They deal only with a \emph{four-dimensional} bulk geometry, but there should be no difficulty in adapting to the five-dimensional case. (In the case of a four-dimensional bulk, one must use the duality of a system of $N_c$ M2-branes with a boundary theory defined on a three-dimensional spacetime. See \cite{kn:ABJM,kn:AdS4}.)}.

This need has been underlined by the very remarkable recent observations made by the STAR collaboration at the RHIC facility, who have reported indirect but convincing evidence \cite{kn:STARcoll,kn:STARcoll2,kn:STARcoll3,kn:STARcoll4} of local rotational motion (``\emph{vorticity}'') in the QGP produced in peripheral collisions of gold nuclei at various impact energies. (The vorticities are deduced from observed polarizations of $\Lambda$/$\overline{\Lambda}$ hyperons: see \cite{kn:STARcoll} for a clear discussion of the experiment and of the discovery.)

Now these ``vortical plasmas'' are extremely complex systems, and one might well be pessimistic as to the prospects for establishing a comprehensive duality between them the AdS$_5$-Kerr black hole, characterized as it is by a very small number of parameters. In fact, even aside from any holographic interpretation, rotating quantum-chromodynamic systems are not simple to describe or even to define, and this in itself is a matter of current research: see for example \cite{kn:yamamoto}, where a lattice approach is found useful, and \cite{kn:fukushima}, where the powerful analogy between rotation and the effects of magnetic fields is exploited.

On the holographic side, the most basic objection to such an enterprise is based on the fact that the spatial sections of the conformal boundary are not flat, as they are in most other applications of gauge-gravity duality (see however Chapter 14 of \cite{kn:nat}). This means that we must take care to ensure that the local spatial curvature on the boundary is always negligible: see the end of Section 2 below. More substantive objections are that the model does not take into account the hydrodynamic aspects of the plasma, as in the celebrated work of Kovtun, Policastro, Son, and Starinets \cite{kn:PSS1,kn:PSS2,kn:KSS,kn:SS}, nor does it allow for the explosive expansion of the plasma \cite{kn:chesler}. The problem is that the AdS$_5$-Kerr metric itself is already a formidably complex object: allowing the geometry to be dynamic will certainly be necessary if a truly realistic model of the vortical plasma is to be constructed, but this is a project for the future.

Here, as an initial step, we set ourselves a much more modest task: can we show that the boundary matter dual to the AdS$_5$-Kerr black hole behaves in a manner that is even moderately realistic when compared with the actual vortical QGP? In particular, can we construct a holographic model which predicts, at least up to order of magnitude, the main parameter reported in \cite{kn:STARcoll}, the \emph{average vorticity} $\omega$, given there as $\omega \approx 9\,\pm 1\,\times 10^{21}\,\cdot\,$s$^{-1}$?

We will argue that this can indeed be done: within the (admittedly large) uncertainties, the AdS$_5$-Kerr model does predict values for the (suitably interpreted) average (over the volume of the plasma) vorticity, at each impact energy, which agree with those reported by the STAR collaboration. (The agreement is good at high impact energies, less good at lower impact energies.) The predicted average over impact energies is $5.3\,\times 10^{21}\,\cdot\,$s$^{-1}$, which, again, in view of the large uncertainties on both the observational and theoretical sides, is acceptable.

Let us proceed to review the AdS$_5$ spacetime and its conformal boundary.

\addtocounter{section}{1}
\section* {\large{\textsf{2. The AdS$_5$-Kerr Geometry: Bulk and Boundary}}}
The AdS$_5$-Kerr metrics were given in \cite{kn:hawk} (but see also \cite{kn:cognola} and \cite{kn:gibperry} for important discussions of the formulae for the physical mass and angular momentum), in the case where the black hole is electrically and magnetically uncharged\footnote{In principle we can also endow the black hole with electric charge, as in \cite{kn:cvet}, in order to model a non-zero baryonic chemical potential. However, just as the metric at conformal infinity for the four-dimensional AdS-Kerr-Newman metric is formally independent of the electric charge, so also the inclusion of electric charge in this case will not modify the form of the boundary metric given in equation (\ref{IJ}) below (which is all we need in this work). Note in this connection that the rate of fall-off of the electric contribution to the metric is larger in five than in four dimensions.}. In five dimensions, the black hole can rotate around two distinct axes simultaneously, so in general one has a pair of rotation parameters, $(a,b)$; but here, for simplicity, we set the second rotation parameter, $b$, equal to zero. The AdS$_5$-Kerr metric in this simplified case takes the form
\begin{flalign}\label{A}
g\left(\m{AdSK}_5^{(a,0)}\right) = &- {\Delta_r \over \rho^2}\Bigg[\,\m{d}t \; - \; {a \over \Xi}\m{sin}^2\theta \,\m{d}\phi\Bigg]^2\;+\;{\rho^2 \over \Delta_r}\m{d}r^2\;+\;{\rho^2 \over \Delta_{\theta}}\m{d}\theta^2 \\ \notag \,\,\,\,&+\;{\m{sin}^2\theta \,\Delta_{\theta} \over \rho^2}\Bigg[a\,\m{d}t \; - \;{r^2\,+\,a^2 \over \Xi}\,\m{d}\phi\Bigg]^2 \;+\;r^2\cos^2\theta \,\m{d}\psi^2 ,
\end{flalign}
where
\begin{eqnarray}\label{B}
\rho^2& = & r^2\;+\;a^2\m{cos}^2\theta, \nonumber\\
\Delta_r & = & (r^2+a^2)\Big(1 + {r^2\over L^2}\Big) - 2M,\nonumber\\
\Delta_{\theta}& = & 1 - {a^2\over L^2} \, \m{cos}^2\theta, \nonumber\\
\Xi & = & 1 - {a^2\over L^2}.
\end{eqnarray}
Here $L$ is the asymptotic AdS curvature length scale, $t$ and $r$ are as usual, and the angular coordinates $\theta, \phi, \psi$ on the topological three-sphere will be described in detail below. The parameters $a$ and $M$ should be regarded\footnote{In the system of natural units we use here, $a$ has units of length, while $M$ has units of squared length.} strictly as quantities describing the geometry of the spacetime; they are related (\emph{but by no means equal}) respectively to the black hole specific angular momentum (angular momentum per unit physical mass) $\mathcal{A}$ and the physical mass\footnote{In natural units, $\mathcal{A}$ has units of length, $m$ of inverse length.} $m$.

In fact, setting $b = 0$ in the formulae given in \cite{kn:gibperry}, we have (if $j$ denotes the black hole's physical angular momentum)
\begin{equation}\label{C}
m\;=\;{\pi M \left(2 + \Xi\right)\over 4\,\ell_{\mathcal{B}}^3\,\Xi^2}, \;\;\;\;\;j\;=\;{\pi M a\over 2\,\ell_{\mathcal{B}}^3\,\Xi^2},
\end{equation}
where $\ell_{\mathcal{B}}$ is the gravitational length scale in the bulk (which is unrelated to the Planck length in physical spacetime).

One should note here that there are several possible distinct definitions of the mass of the black hole; this is discussed in detail in \cite{kn:gibperry}. In particular, in five dimensions one has to decide whether to include the contribution of the Casimir energy at infinity. This is appropriate in some applications (certainly if one is interested in the holography of the conformal anomaly \cite{kn:awad,kn:papaskend}), but not in the application to the physical QGP, which we are attempting to describe here. Therefore we follow \cite{kn:gibperry} (see in particular their Footnote 5), where the constant of integration arising in the First Law is systematically set equal to zero, so that empty AdS$_5$ has zero mass, and there is no Casimir contribution to the energy density of the boundary field theory.

From the equations (\ref{C}) we see that the angular momentum to (physical) mass ratio, or specific angular momentum, is given by
\begin{equation}\label{CC}
\mathcal{A}\;=\;{2 a \over 2 + \Xi}\;=\;{2 a \over 3 - \left(a^2/L^2\right)}.
\end{equation}
Note carefully that this differs from the four-dimensional case, where $a$ itself is the specific angular momentum.

Other important characteristics of the black hole can be computed from its geometry: for example, the Hawking temperature is given \cite{kn:gibperry} by
\begin{equation}\label{CCC}
T\;=\;{r_H\left(1 + {r_H^2\over L^2}\right)\over 2\pi \left(r_H^2 + a^2\right)} + {r_H\over 2\pi L^2},
\end{equation}
where $r_H$ denotes the horizon radius (which can be regarded as a function of $M$ and $a$ through its definition as the largest root of $\Delta_r$), and the entropy by
\begin{equation}\label{CCCC}
S\;=\;{\pi^2\left(r_H^2 + a^2\right)r_H\over 2\ell_{\mathcal{B}}^3\Xi}.
\end{equation}

Finally, we note that the angular coordinates $\theta, \phi, \psi$ used in \cite{kn:hawk}, which we follow here, are not the familiar polar coordinates on the three-sphere: for example, one sees that, when $\mathcal{A} = 0$, the angular part of the metric is $r^2\left(\m{d}\theta^2\,+\,\sin^2\theta\m{d}\phi^2\,+\,\cos^2\theta\m{d}\psi^2\right)$, which is indeed the usual round metric on $S^3$ with radius $r$, but not in polar coordinates. Instead, these are the coordinates\footnote{In terms of the usual coordinates ($x_0, x_1, x_2, x_3$) on Euclidean $\bbr^4$, we have, for a three-sphere of radius $r$,
$$x_0 = r\cos\psi \cos\theta$$
$$x_1 = r\sin\psi \cos\theta$$
$$x_2 = r\cos\phi \sin\theta$$
$$x_3 = r\sin\phi \sin\theta.$$
Notice however that, if we set $\psi = 0,$ then the remaining coordinates can be interpreted as polar coordinates on a hemisphere of $S^2$.}
usually used to describe $S^3$ when it is regarded as a principal $U(1)$-bundle over $S^2$, that is, as the Hopf bundle \cite{kn:steen}. The two coordinates $\phi$ and $\psi$ both run from $0$ to $2\pi$, while $\theta$ runs from $0$ to $\pi/2$; thus in fact each fixed value of $\theta$ corresponds to a two-torus, \emph{except} in the ``degenerate'' cases $\theta = 0, \pi/2$, which correspond to one-dimensional circles. In particular, the single condition $\theta = \pi/2$ reduces the dimensionality by \emph{two}: it means that we are on the equator of $S^3$, described by a single angular coordinate, $\phi$.

The geometry of the conformal boundary is fixed by means of a conformal re-scaling of the metric $g(\m{AdSK}_5^{(a,0)})$ as the limit is taken to infinity. We take the boundary metric to be
\begin{equation}\label{IJ}
g\left(\m{AdSK}_5^{(a,0)}\right)_{\infty}\;=\;-\,\m{d}t^2 \;+\;{2a\,\m{sin}^2\theta\,\m{d}t \m{d}\phi\over \Xi} \;+\; {L^2 \, \m{d}\theta^2 \over 1 - (a/L)^2\m{cos}^2\theta} \;+\; {L^2 \m{sin}^2\theta\m{d}\phi^2\over \Xi}\;+\;L^2\cos^2\theta \,\m{d}\psi^2,
\end{equation}
obtained from $g(\m{AdSK}_5^{(a,0)})$ by extracting a conformal factor $r^2/L^2$, taking the limit $r \rightarrow \infty$, and then doing some algebraic simplifications. We have chosen the conformal factor so that the time coordinate $t$ represents proper time for a stationary observer at infinity located at one of the poles ($\theta = \psi = 0$); we can take this observer to be an outside observer, fixed in the laboratory. In each case we consider below (particles with zero, respectively non-zero angular momentum), $\m{d}\phi/\m{d}t$ represents an angular velocity as measured by this observer\footnote{In view of the ``large'' angular velocities arising in our application, it is natural to ask whether we should give here a relativistic account of angular velocity. The answer is that, while the angular velocities here seem large by ordinary standards, they are in fact surprisingly \emph{small} in view of the size of the systems in question. Perhaps the best way to see this is to use natural units, taking the femtometre as the basic unit. In these units, we will find later that a typical angular velocity here is on the order of $0.004 \;\m{fm}^{-1}$. For systems with a radius of a few femtometres, this does not lead to relativistic velocities, so we will not take relativistic effects into account here. See \cite{kn:newbes} for a discussion of this. Another way of seeing the point is to note that, during the lifetime of the plasma, a system rotating at such angular velocities executes far less than one complete revolution \cite{kn:romat}.}.

We stress, because it will be crucial later, that the physical parameters of the AdS$_5$-Kerr black hole (the physical mass $m$, and the specific angular momentum $\mathcal{A}$) are related to the geometric parameters $M$ and $a$ in a surprisingly indirect manner (equations (\ref{C}) and (\ref{CC}) above). This is due to the ubiquitous presence of the quantity $\Xi$, which appears both in the bulk metric and in its counterpart on the conformal boundary. This in turn is required in order to maintain the regularity of the geometry in both cases\footnote{This is particularly clear in the boundary geometry: there, the circumference of a circle of the form $\theta = \theta_0 = $ constant, located on the two-dimensional hemisphere $\psi = 0$, is $2\pi L\sin\theta_0/\sqrt{\Xi}$, while its radius, measured from the pole (of both the 3-sphere and the two-dimensional hemisphere) along the hemisphere, is $L\,\int_0^{\theta_0}{d\theta\over \sqrt{1 - (a^2/L^2)\cos^2\theta}}$; the ratio only tends to $2\pi$ as $\theta_0 \rightarrow 0$ because the $\sqrt{\Xi}$ factor is present.}.

Before proceeding, we draw the reader's attention to the following (ultimately) very important aspect of the AdS$_5$-Kerr geometry. Examining the metric, we see that, if no constraint is imposed on $a$, then the signature of the coefficient of d$\theta^2$, $1/\Delta_{\theta}$, can vary, depending on $\theta$: it is positive for $\theta > \arccos\left(L/a\right)$, but \emph{negative} for $\theta < \arccos\left(L/a\right)$. At $\theta = \arccos\left(L/a\right)$ there is a ``singularity'', which admittedly proves to be a coordinate ``singularity''; nevertheless, it seems clear that this kind of behaviour is a complication which we should consider only if it cannot be avoided. We therefore require, in this work, that $a$ \emph{should always be smaller than} $L$.

In fact, it is natural to impose a stronger condition on $a$, for the following reason. The status of cosmic censorship in higher-dimensional spacetimes is currently unsettled: see for example \cite{kn:horsant,kn:binwang} and references therein. In discussing gauge-gravity duality, however, we clearly need to assume that some form of censorship does hold, since the boundary theory (representing an equilibrated plasma) certainly does have a well-defined temperature and entropy density; and so the dual object in the bulk must have a well-defined event horizon. For the geometry we are discussing here, censorship takes the form $2M\,\geq a^2$, or, expressed in terms of the \emph{physical} mass,
\begin{equation}\label{CA}
8\,\ell_{\mathcal{B}}^3\,m\,\Xi^2\;\geq \;\pi\,a^2\,\left(2\,+\,\Xi \right);
\end{equation}
we see from this that, if $m$ takes any fixed finite value, then (because of the factor of $\Xi^2$ on the left) this condition will be violated, for some $a$ strictly smaller than $L$, if we steadily increase $a$ from zero towards $L$. With these assumptions, then, $a$ cannot even come arbitrarily close to $L$, much less attain that value. (See \cite{kn:mcong} for a discussion of this in the four-dimensional case; see also \cite{kn:gwak}.) This is a useful piece of information, for it means that, in all of the many expressions depending on the reciprocal of $\Xi$ (or its square root), we are not dealing with quantities which can be arbitrarily large.

The physics of the AdS$_5$-Kerr black hole determines, according to the gauge-gravity duality \cite{kn:nat}, that of a field theory on the boundary ($r\,\rightarrow\,\infty$); this field theory is held to approximate, to some extent, to the QGP. Thus, the Hawking temperature $T$ of the black hole corresponds to the temperature of the plasma, the ratio of the black hole's entropy to its (physical) mass is equal to the ratio of the plasma entropy density $s$ to its energy density $\varepsilon$, and similarly the black hole specific angular momentum parameter $\mathcal{A}$ is interpreted as the ratio of the QGP angular momentum density $\alpha$ to its energy density. Finally, the bulk curvature length scale is in principle given a dual interpretation on the boundary by the ``holographic dictionary''; it is related to the number of colours and the 't Hooft coupling of the boundary field theory\footnote{In detail, one has
\begin{equation}\label{DD}
{\ell_{\mathcal{B}}^3\over L^3} = {\pi\over 2N_c^2},\;\;\;\;\;\;\;{\ell_s^4\over L^4}\;=\;{1\over \lambda},
\end{equation}
where $N_c$ is the number of colours in the boundary field theory, $\lambda$ is the 't Hooft coupling in that theory, and $\ell_s$ is the string length scale. The duality is useful only when the bulk can be treated classically and when strings can be treated as point particles: that is, when $L$ is large relative to the other two length scales. These conditions are certainly satisfied here, with the lower bound on $L$ we are about to describe.}. In practice, $L$ is not known, but its value can be usefully constrained, for example as follows.

Notice first that $L$ cannot be scaled away here, since the event horizon is topologically spherical. This is not a new observation (see \cite{kn:nat}, Chapter 14, for a detailed discussion in the non-rotating case), but it is important, because it is related to the fact that the boundary field theory is defined on a compact space, whereas of course the actual space which the plasma inhabits is not compact. In principle, this leads to various properties which may not be welcome or realistic (for example, perturbation spectra of fields on the black hole background become discrete, there is a phase transition (the Hawking-Page transition) which may not however correspond well with the actual phase transition (hadronization) experienced by the plasma, and so on). It follows that the use of the topologically spherical Kerr geometry in holographic models (as in \cite{kn:sonner,kn:schalm} and here) is really only acceptable if the volume of the compact space is very large relative to the size of the system being studied. As we now explain in detail, in our case this is in fact implied by our discussion above.

The spatial manifolds defined by $t = \tau = $ constant on the boundary have the geometry of a deformed three-sphere, with metric
\begin{equation}\label{IA}
g\left(\m{AdSK}_5^{(a,0)}\right)_{\infty}(t = \tau)\;=\; {L^2 \, \m{d}\theta^2 \over 1 - (a/L)^2\m{cos}^2\theta} \;+\; {L^2 \m{sin}^2\theta\m{d}\phi^2\over \Xi}\;+\;L^2\cos^2\theta \,\m{d}\psi^2.
\end{equation}
The volume of the three-sphere with this metric is given by
\begin{equation}\label{IB}
V(L, \mathcal{A})\;=\;2\pi^2L^3\left[{2L^2\over a^2}\,\left({1\over \sqrt{\Xi}}\,-\,1\right)\right],
\end{equation}
where, by inverting equation (\ref{CC}), we regard $a$ as a function of $\mathcal{A}$ and $L$ (and hence we can do the same for $\Xi$, the last member of (\ref{B})). We have written the volume in this form so as to enable a comparison with the volume of the sphere in the case where $\mathcal{A} \rightarrow 0$, which is of course $2\pi^2L^3$.

Now, our condition that $a$ should be strictly smaller than $L$ implies that $\Xi$ must be positive, so it is clear from equation (\ref{CC}) that $\mathcal{A}$ is always smaller than $a$, and therefore than $L$. In short, we always have
\begin{equation}\label{D}
\mathcal{A}\;<\; L.
\end{equation}
Since $a$ is smaller than $L$, the expression in square brackets in equation (\ref{IB}) is larger than unity for all non-zero $\mathcal{A}$; so if we use $2\pi^2L^3$ to compute the volume, we will under-estimate it (by, as it will turn out, a factor of about 3). On the other hand, for collisions of gold ions at 200 GeV impact energy per pair and 20$\%$ centrality, we find below that a reasonable estimate of the ratio of the plasma angular momentum density to its energy density is $\approx$ 72 femtometres, and in the holographic model this is $\mathcal{A}$; so $L$ should be larger than this. In this way we obtain an extremely conservative lower bound for the spatial volume in this case: $V(L, \mathcal{A}) \geq \approx 7.4 \times 10^6$ fm$^3$; the real figure is probably an order of magnitude larger. At lower impact energies, the volume computed in this way is somewhat smaller, but, in all cases where vorticity has actually been detected experimentally, it is essentially infinite compared to the size of the system we are treating. (The region occupied by the equilibrated plasma resulting from a collision at this centrality has a total volume on the order of 100 fm$^3$.) We do not therefore anticipate any difficulties on this score.

We now turn to our main problem: can we compute the angular velocity of the matter at infinity which is dual to an AdS$_5$-Kerr black hole with a specified ratio $\mathcal{A}$ of angular momentum to mass? That is, can we give a holographic estimate of the vorticity of a ``QGP-like'' fluid in terms of the ratio $\alpha/\varepsilon$ of the plasma's angular momentum density to its energy density?

\addtocounter{section}{1}
\section* {\large{\textsf{3. The AdS$_5$-Kerr Geometry: Angular Velocity at Infinity}}}
Our task now is to use the AdS$_5$ geometry to estimate angular velocities on the boundary. This is not quite straightforward.

The black hole introduced above, with an angular momentum per unit mass value equal to $\mathcal{A}$, is dual to a fluid on the boundary, described by a field theory in the usual holographic manner. Since the duality is a complete equivalence, we assume that this single parameter, $\mathcal{A}$, sets the scale for \emph{all} rotational phenomena in the boundary theory, as it clearly does in the bulk. This paucity of parameters is an indirect consequence of the ``no-hair'' theorems, which indeed dictate that the bulk geometry is determined by a very small number of physical quantities, including (in our case) a single angular momentum parameter\footnote{In the higher-dimensional context, ``no-hair'' statements apply when one fixes the topology of the event horizon to be spherical, as we are doing here. See \cite{kn:ida}.}.

This means that we are committed to a matter model on the boundary in which there is only one angular momentum scale, which of course the holographic correspondence dictates should be equal to $\mathcal{A}$. Similarly there is only one angular velocity scale; the boundary matter rotates like a rigid body.

The detailed way in which vorticity develops in the actual plasma is very complex. The vorticity is thought to be initially focused mainly in a thin layer, with a thickness presumably measured in fractions of a femtometre, between the participant and spectator matter \cite{kn:ivansold1,kn:ivansold2}; it then propagates inward, through viscous effects.

The motion of this system cannot, of course, be literally pictured as a simple rotating object. When \cite{kn:STARcoll} characterizes this system by an angular velocity given as $9\,\pm 1\,\times 10^{21}\,\cdot\,$s$^{-1}$ (with a certain systematic uncertainty), the intention is not, of course, to claim that the matter in any actual QGP vortex rotates at this rate. Instead, this quantity is intended to give an overall, \emph{averaged} (over the volume of the plasma, over the full range of impact energies, and so on) indication of the internal motion of the vortical plasma. One should in fact interpret this number as \emph{nothing more than a measure of the extent to which the internal dynamics of the vortical plasma differs from that of  the plasmas produced in central collisions}. Our task here is to try to use a computation of the angular velocity at infinity for the AdS$_5$-Kerr spacetime to \emph{reproduce this number} (and the allied numbers which lead to it), not to provide a detailed holographic model of the velocity field in the actual plasma. That would be an interesting and important project, but it is far beyond our ambitions here.

In view of all this, we propose to compute this average vorticity by examining massive particles in the boundary theory, with angular momentum per unit mass equal to $\mathcal{A}$, having equatorial orbits. The boundary metric, given in equation (\ref{IJ}), has a Killing vector field proportional to $\partial_{\phi}$ (the normalization being fixed by our assumption that $\phi$ has periodicity $2\pi$). The inner product of this vector field with the unit tangent to the particle worldline, $\dot{t}\,\partial_{\,t} + \dot{\phi}\,\partial_{\phi}$ (where dots denote differentiation with respect to the proper time of the particle), gives us the angular momentum per unit mass.

Let us begin by considering such particles with zero angular momentum. Computing the inner product as above, we have in that case
\begin{equation}\label{JO}
{\dot{t}a\over \Xi}\;+\;{\dot{\phi}L^2\over \Xi}\;=\;0,
\end{equation}
from which we have at once, denoting the angular velocity of these particles by $\omega_0$,
\begin{equation}\label{JK}
\omega_{0}\;=\;{-\,a\over L^2}.
\end{equation}
Because of this equation, it is sometimes said that ``the boundary rotates with angular velocity $-\,a/L^2$''. But this is just a way of describing the motion of particles with a distinguished angular momentum to mass ratio, namely zero. Here however we are not interested in such particles: we wish the particles to have an angular momentum to mass ratio $\mathcal{A}$, corresponding holographically to the bulk black hole with angular momentum to mass ratio having that same value.

We propose that the physical angular velocity $\omega$, corresponding to the vorticity of the plasma as observed experimentally, should be computed as the difference between the angular velocity of particles with non-zero angular momentum, $\omega_{\mathcal{A}}$, and that of fictitious particles with zero angular momentum, $\omega_{0}$: we have $\omega\,=\,\omega_{\mathcal{A}}\,-\,\omega_{0}$. (To put it more picturesquely: we compute the angular velocity by using a frame at infinity which ``co-rotates with the boundary''.)

To compute $\omega_{\mathcal{A}}$, we have again, from equation (\ref{IJ}),
\begin{equation}\label{JJ}
{\dot{t}a\over \Xi}\;+\;{\dot{\phi}L^2\over \Xi}\;=\;\mathcal{A}.
\end{equation}
We need to supplement this with the fact that $\dot{t}\partial_t + \dot{\phi}\,\partial_{\phi}$ is a unit vector in the boundary geometry, that is,
\begin{equation}\label{JJJ}
-\,\dot{t}^{\,2}\;+\;{2\,\dot{t}\dot{\phi}\,a\over \Xi}\;+\;{\dot{\phi}^{\,2}L^2\over \Xi}\;=\;-\,1.
\end{equation}
Using this equation, we can eliminate $\dot{t}$, so obtaining a quadratic equation for $\omega_\mathcal{A}$:
\begin{equation}\label{JJJJ}
L^2\,\omega_\mathcal{A}^2\;+\;2\,a\,\omega_\mathcal{A}\;+\;{a^2\over L^2}\,\left({1\;-\;{\mathcal{A}^2\over a^2}\,\Xi^2\over 1\;+\;{\mathcal{A}^2\over L^2}\,\Xi}\right)\;=\;0.
\end{equation}
Solving this and subtracting $\omega_0$ as suggested, we obtain two equal and oppositely signed values for $\omega$, corresponding of course to the two possible directions of rotation; taking the positive value we have finally,
\begin{equation}\label{K}
\omega\;=\;{\mathcal{A}\over L^2}\,\sqrt{{\Xi\over 1\;+\;{\mathcal{A}^2\over L^2}\,\Xi}},
\end{equation}
where, through equation (\ref{CC}), we can regard $\Xi$ as the function of $\mathcal{A}$ given by
\begin{equation}\label{KA}
\Xi\;=\;{\sqrt{1\,+\,{3\mathcal{A}^2\over L^2}}\;-\;1\;-\;{\mathcal{A}^2\over L^2}\over {\mathcal{A}^2\over 2L^2}} .
\end{equation}

These are the relations we seek: $\omega$ is interpreted as the vorticity of the plasma, and, if that plasma has angular momentum density $\alpha$ and energy density $\varepsilon$, we have $\alpha/\varepsilon = \mathcal{A}$. The vorticity of the plasma is now computed in terms of its physical parameters (together with $L$, see below.)

The relation between $\omega$ and $\mathcal{A}$ appears to be unusual: note for example that $\omega$ is small when $\mathcal{A}$ is sufficiently large. Of course, unusual relations between angular velocity and angular momentum are not unexpected for matter associated with rotating black holes: ``frame dragging'' has the same origin. In this specific case, one way to understand the structure of equation (\ref{K}) is to observe that it ensures that causality is never violated here. To see this, let us compute the linear velocity of objects on the equator of the boundary sphere, having angular velocity $\omega$. From equation (\ref{IJ}) we see that the circumference is $2\pi L/\sqrt{\Xi}$, which is larger than $2\pi L$, by a factor which \emph{diverges} as $a$ approaches $L$ ---$\,$ so indeed large values of $\mathcal{A}$ (and $a$) might easily give rise to very high linear velocities in this case. For an object with angular velocity $\omega$, the linear velocity of this potentially large circumference is given by
\begin{equation}\label{KK}
v_{\omega}\;=\;{\omega L\over \sqrt{\Xi}},
\end{equation}
and this could exceed unity, violating causality, as $a$ and $\mathcal{A}$ approach $L$, \emph{unless} the dependence of $\omega$ on $a$ causes it to tend to zero sufficiently rapidly in the same limit\footnote{Of course, we know that (as long as the physical mass of the black hole remains finite) $a$ cannot actually come arbitrarily close to $L$ without violating cosmic censorship. We cannot rely on this to avoid causality violation here, however, for the following reason. Censorship is expressed for this black hole by the inequality (\ref{CA}), which involves, on the left, the quantity $\ell_{\mathcal{B}}^3$. Unfortunately we do not know this quantity (apart from the fact that it must be small relative to $L^3$, see the holographic ``dictionary'', above), so in practice we cannot say precisely how large $a$ can be relative to $L$.}. In short, the angular velocity \emph{must} decrease towards zero when the parameter $a$ is sufficiently large, in order to avoid violating causality.

In fact, substituting equation (\ref{K}) into (\ref{KK}) we find
\begin{equation}\label{KKK}
v_{\omega}\;=\;{\mathcal{A}\over L}\,\sqrt{{1\over 1\;+\;{\mathcal{A}^2\over L^2}\,\Xi}},
\end{equation}
and (in view of inequality (\ref{D}) and the fact that $\Xi$ is always positive here) this does always satisfy causality, because the factor of $\sqrt{\Xi}$ in the denominator of the right side of (\ref{KK}) has been cancelled.

In principle, equations (\ref{K}) and (\ref{KA}) allow us to make a holographic prediction of typical angular velocities characterizing a vortical plasma in terms of the ratio of its angular momentum density $\alpha$ to its energy density $\varepsilon$. In practice, unfortunately, this is not possible: for it is clear that the relation is mediated by $L$, which we do not know. As in the computation of the spatial volume at infinity, we have to find a way of circumventing this.

\addtocounter{section}{1}
\section* {\large{\textsf{4. The Bound on $\omega$, and a Conjecture }}}
The black hole parameter $\mathcal{A}$ is computed holographically from the ratio $\alpha/\varepsilon$, which can be estimated explicitly from physical data and phenomenological models, and we will explain how to do this shortly. Fixing $\mathcal{A}$ at a definite value in this manner, we may use equations (\ref{K}) and (\ref{KA}) to regard $\omega$ formally as a function of $L$, $\omega(L)$. We now ask: what form does this function take?

We stress here that, in doing so, we are not ``varying'' $L$; we are merely exploring whether it is possible to deduce anything regarding $\omega$ without knowing what value $L$ actually takes.

We know (from the inequality (\ref{D})) the domain on which $L$ is defined: it is the open interval $(\mathcal{A},\;\infty)$. Clearly $\omega$ vanishes as $L \rightarrow \infty$, as expected, since $\omega \approx \mathcal{A}/L^2$ in that limit: this is like a system with a large moment of inertia. We have seen that it \emph{also} vanishes as $L$ tends down to $\mathcal{A}$, for reasons connected with causality. Now since $\omega$ is continuous and positive as a function of $L$, \emph{it follows that $\omega$ is bounded above}. This means that it is possible at least to put an ($\mathcal{A}$-dependent) bound on $\omega$ without knowing $L$.

In fact, $\omega(L)$ can be completely described analytically; the analysis is rather intricate but of course essentially elementary. One finds\footnote{One regards $\omega \mathcal{A}$ as a function of the dimensionless variable $\sigma = L/\mathcal{A}$, computes the derivative and sets it equal to zero; a somewhat elaborate algebraic manipulation reduces this condition to solving the octic $24\sigma^8 + 63\sigma^6 - 64\sigma^4 - 176\sigma^2 + 48 = 0$. This is a quartic in $\sigma^2$ and so it can be solved explicitly and exactly if desired. One finds that there are two real positive solutions for $\sigma$, of which however only one, given approximately by $\sigma \;\approx \; 1.2499$, allows the crucial inequality (\ref{D}) to be satisfied. Substituting this value of $\sigma$ back into equations (\ref{K}) and (\ref{KA}), one obtains the maximum possible value of $\omega$, expressed as a multiple of $1/\mathcal{A}$.} that this function has a unique maximum, $\omega_{\m{max}}$, given by
\begin{equation}\label{M}
\omega_{\m{max}}\;=\;{\varkappa\over \mathcal{A}}\;\approx \; {0.2782\over \mathcal{A}},
\end{equation}
where $\varkappa$ is a pure number which can be computed to any desired precision, having the indicated approximate value. Clearly, \emph{no matter what value} $L$ actually takes, $\omega$ can never exceed this value, once $\mathcal{A}$ has been fixed.

Notice that, even when the particles in the model with angular velocity $\omega$ have the maximal possible angular velocity for a given value of $\mathcal{A}$, they move at a velocity $v_{\omega}(\omega_{\m{max}}),$ computed using equation (\ref{KKK}), from which we obtain
\begin{equation}\label{MM}
v_{\omega}(\omega_{\m{max}})\;\approx \; 0.750,
\end{equation}
that is, well below light speed. Thus, while we saw earlier that our model is always consistent with causality, the latter does \emph{not} explain the existence of this maximal angular velocity for given $\mathcal{A}$ ---$\,$ an important point\footnote{Of course, in general, the question of causality can indeed be crucial in discussions of (sufficiently large) rapidly rotating systems: see for example \cite{kn:chern} and references therein. The point is that this is not the case here.}.

In terms of the physical quantities, we have the bound
\begin{equation}\label{ALPHA}
 \omega\;\leq\;\varkappa\;{\varepsilon\over \alpha}\;\approx \; 0.2782\;{\varepsilon\over \alpha}.
\end{equation}
This is of interest for three reasons. First, of course, we can evaluate the right side using phenomenological models, and compare it with the reported data on vorticity; this gives us a new way of testing holography in general. Again we stress that peripheral collisions are \emph{generic}: if holography fails this test, then its general validity will be called into question.

Secondly, this inequality identifies for us the ``right'' variable to be examined in studies of the vortical plasma, namely $\varepsilon/\alpha$. Just as realistic, astrophysical black holes are conventionally described by their angular momentum to mass ratio, so here holography identifies (the \emph{reciprocal} of) that ratio as the one on which we should focus attention. Notice in this connection that $\alpha$ and $\varepsilon$, as densities of conserved quantities, can be expected to vary strongly as the plasma expands; but their ratio is expected to be much more stable.

Finally, and related to this second point: the inequality focuses our attention on a specific value of $\omega$ for each peripheral collision, namely $0.2782\;\varepsilon/\alpha$. \emph{This specific value is now strongly distinguished}, and we should expect it to play some special role. In fact, we can speculate that it appears here as the right side of an \emph{inequality} merely as an artefact of our ignorance of $L$. Perhaps we should expect that $0.2782\;\varepsilon/\alpha$ is close to the \emph{actual} value of the vorticity, for impact parameters which are not very small or large (see below). In short, we conjecture that the vorticity is actually given holographically as
\begin{equation}\label{BETA}
 \omega\;=\;\varkappa\;{\varepsilon\over \alpha}\;\approx \; 0.2782\;{\varepsilon\over \alpha}
\end{equation}
for a range of impact parameters or centralities to be described below.

This conjecture can now be tested.

\addtocounter{section}{1}
\section* {\large{\textsf{5. The Data}}}
Before we begin, we should be open regarding the fact that precision is not to be looked for in these computations. As is well known, the field theories considered in the gauge-gravity duality are not (in all regimes) closely similar to QCD: there are important analogies (discussed very clearly in \cite{kn:nat}) but there are basic differences. Furthermore, as we discussed, the no-hair theorems strongly restrict the number of parameters available to describe the dual plasma: we have only one angular momentum parameter, but $\Lambda$/$\overline{\Lambda}$ hyperon polarization has a longitudinal component apart from the component we study here; the model cannot describe this. The observational data, too, suffer from large uncertainties: \cite{kn:STARcoll} reports a large systematic uncertainty in the given values of the polarization percentages. (The principal difficulty here is that, to relate vorticities, which are not directly observable, to polarizations, one needs to know the precise temperature corresponding to a given energy density; and this is a notoriously difficult problem \cite{kn:bus}.)

Rather than continuously repeat these points, we will simply proceed to a detailed comparison with the data; our main priority, however, is always to assure ourselves that (\ref{ALPHA}) and (\ref{BETA}) impose constraints on or predict values for hyperon polarizations that are at least of a reasonable order of magnitude. Our secondary objective is to explain clearly how the relevant calculations can be done when more precise data, and perhaps more sophisticated holographic models, become available in future.

We need estimates for the angular momentum of the QGP in peripheral collisions, and also for the energy density, for various impact energies and centralities. For the former, we rely on \cite{kn:jiang}, where estimates\footnote{Note that, in \cite{kn:jiang}, the convention is used in which $\omega$ is twice as large as in \cite{kn:STARcoll} and here.} using the AMPT (``A Multi-Phase Transport'') model are given; for the latter we use \cite{kn:sahoo}, where very detailed computations of many relevant parameters (using a colour string percolation model) are given. In both cases, the values given seem reasonable. Other models might be chosen, but the differences are unlikely to be sufficiently large as to modify our general conclusions.

Proceeding in this way, we can (with simple additional assumptions regarding the geometry of the overlapping nuclei) compute the right side of the inequality (\ref{ALPHA}), and thus place an explicit bound on the vorticity in each case. We can relate this to the average polarizations of primary $\Lambda$ and $\overline{\Lambda}$ hyperons by using the same equation\footnote{The primes indicate that these are the polarizations for ``primary'' hyperons, which means that we are neglecting the ``feed-down'' effect; see \cite{kn:hyper} for the theory of this, and \cite{kn:STARcoll} for a discussion of its effect on the STAR observations.} as in \cite{kn:STARcoll},
\begin{equation}\label{P}
\overline{\mathcal{P}}_{\Lambda'}\,+\,\overline{\mathcal{P}}_{\overline{\Lambda}'}\;=\;{\omega\over T},
\end{equation}
where natural units are used and $T$ is the temperature as usual. (Actually, we will use it ``in reverse'': in \cite{kn:STARcoll}, it is used to deduce vorticities from polarizations, here we do the opposite.) This equation allows us to express the vorticity bound as a bound on the total polarization, obtaining inequalities of the form
\begin{equation}\label{PP}
\left[\overline{\mathcal{P}}_{\Lambda'}\,+\,\overline{\mathcal{P}}_{\overline{\Lambda}'}\right]\left(\sqrt{s_{\m{NN}}},\,\mathcal{C}\right) \;\leq \; \Phi(\sqrt{s_{\m{NN}}}, \mathcal{C})\;\equiv\;{\varkappa\, \varepsilon \over \alpha T},
\end{equation}
where $\mathcal{C}$ denotes centrality and where we now know how to compute $\Phi(\sqrt{s_{\m{NN}}}, \mathcal{C})$ from phenomenological estimates in each instance; it is expressed as a percentage.

We consider three different impact energy regimes.

\subsubsection*{{\textsf{5.1 Collisions at 62.4 GeV and Higher}}}
Let us begin by considering the highest-energy collisions studied by the STAR collaboration, particularly those with an impact energy of 200 GeV per pair, for which the observation of vorticity \cite{kn:STARcoll2} is most unambiguous.

In \cite{kn:STARcoll,kn:STARcoll2}, the focus is on collisions with $20\%$ to $50\%$ centrality. This means \cite{kn:bron,kn:olli2} that the impact parameters vary from around 6.75 femtometres (fm) up to around 10.5 fm. On this domain, one finds \cite{kn:jiang} that the angular momentum imparted to the plasma in 200 GeV collisions steadily decreases, from about $110000$ (in natural units; in conventional units, $110000 \cdot \hbar \,$) at $b = 6.75$ fm to around $40000$ at $b = 10.5$ fm. However, the volume of the overlap region \emph{also} decreases as the impact parameter increases, and we find that, to a good approximation, the two effects cancel for collisions with $20\%$ to $50\%$ centrality: that is, the angular momentum density $\alpha$ is roughly independent of $b$ in this range of $b$ values. We therefore focus on collisions at $20\%$ centrality, since, in this range, these collisions are least affected by the variations of the nuclear density near the boundary of the nucleus, and by other effects associated with the very small volumes of the plasma produced by high-centrality collisions.

The volume of the plasma sample can then be computed in an elementary way, in a ``hard sphere'' model, using the formula \cite{kn:wolf} for the volume of the intersection of two spheres. However, as explained in \cite{kn:jiang}, this underestimates the effective volume, both because the ``sharp edge'' assumption is inadequate (a Woods-Saxon profile is used in \cite{kn:jiang}) and because, in reality, some nucleons outside the overlap zone contribute to the fireball, effectively increasing the volume. This effect is estimated in \cite{kn:jiang} to be of order 2 to 3, depending on the impact parameter: in our case it is around 2. In addition, we must of course take into account relativistic contraction, estimated in \cite{kn:phobos} to be roughly 7 for the equilibrated plasma.

Taking all this into account, we find that, for $\sqrt{s_{\m{NN}}} = 200$ GeV collisions at $\mathcal{C} = 20\%$ centrality, the angular momentum density is approximately given by
\begin{equation}\label{N}
\alpha\left(\sqrt{s_{\m{NN}}} = 200\, \m{GeV},\,\mathcal{C} = 20\%\right)\;\approx\; 758 \; \m{fm}^{- 3}.
\end{equation}

According to \cite{kn:sahoo}, the energy density in this case is approximately $10.55/\m{fm}^4$, and so we compute the maximal vorticity, according to equation (\ref{M}), as
\begin{equation}\label{O}
\omega_{\m{max}}\left(\sqrt{s_{\m{NN}}} = 200\, \m{GeV},\,\mathcal{C} = 20\%\right)\;\approx\; 0.00387 \;\m{fm}^{-1}.
\end{equation}
In order to compute a total polarization from this, we need to use the (initial) temperature of the plasma, and this is the point of greatest uncertainty, as mentioned above. Using a temperature of approximately 190 MeV \cite{kn:sahoo} (which may be an over-estimate, so our result may well be somewhat too low), we can express the vorticity bound in this case in the form
\begin{equation}\label{Q}
\left[\overline{\mathcal{P}}_{\Lambda'}\,+\,\overline{\mathcal{P}}_{\overline{\Lambda}'}\right]\left(\sqrt{s_{\m{NN}}} = 200\, \m{GeV},\,\mathcal{C} = 20\%\right) \;\leq \;\approx\, 0.402\%.
\end{equation}
When the first observations of polarization of $\Lambda$ and $\overline{\Lambda}$ hyperons were announced, such a value was too small to be detected (see the rightmost points in Figure 4 of \cite{kn:STARcoll}). Subsequent analysis of a much larger data set \cite{kn:STARcoll2} has however found evidence of such polarization, reporting values of $$\overline{\mathcal{P}}_{\Lambda'}\left(\sqrt{s_{\m{NN}}} = 200\, \m{GeV},\,\mathcal{C} = 20\%\right)\,\approx \, 0.277\,\pm\,0.040\,(+\,0.039\,-0.049)\%$$ and $$\overline{\mathcal{P}}_{\overline{\Lambda}'}\left(\sqrt{s_{\m{NN}}} = 200\, \m{GeV},\,\mathcal{C} = 20\%\right) \,\approx \, 0.240\,\pm\,0.045\,(+\,0.061\,-\,0.045)\%,$$ the uncertainties being statistical and systematic respectively.

This is the most precise vorticity observation thus far reported, and is considered to be particularly trustworthy because the values of $\overline{\mathcal{P}}_{\Lambda'}\left(\sqrt{s_{\m{NN}}} = 200\, \m{GeV},\,\mathcal{C} = 20\%\right)$ and $\overline{\mathcal{P}}_{\overline{\Lambda}'}\left(\sqrt{s_{\m{NN}}} = 200\, \m{GeV},\,\mathcal{C} = 20\%\right)$ are considered (in \cite{kn:STARcoll2}) to be essentially indistinguishable with these uncertainties, and this is expected on theoretical grounds. We see that, by the same measure, these results are also consistent with both with our vorticity bound (\ref{ALPHA}) \emph{and} with our conjectured equality, (\ref{BETA}).

If one repeats this calculation for collisions at an impact energy of 62.4 GeV, one finds that the energy density is of course lower (about $7.59/\m{fm}^4$), as is the temperature (about 179 MeV) and that the angular momentum density also drops, \emph{but more sharply}, to around $236.4/\m{fm}^3$ (it scales approximately linearly with $\sqrt{s_{\m{NN}}}\;$ \cite{kn:jiang}): this pattern is seen throughout these calculations. The result is a much less\footnote{That is, the vorticity is predicted to be larger for smaller angular momentum densities; this is clear from (\ref{ALPHA}) directly, and it is in fact in agreement with all of the reported data. See \cite{kn:jiang} for the physics of this.} stringent bound,
\begin{equation}\label{R}
\left[\overline{\mathcal{P}}_{\Lambda'}\,+\,\overline{\mathcal{P}}_{\overline{\Lambda}'}\right]\left(\sqrt{s_{\m{NN}}} = 62.4\, \m{GeV},\,\mathcal{C} = 20\%\right) \;\leq \;\approx\, 0.980\%,
\end{equation}
which might well be detectable in an analysis similar to that of \cite{kn:STARcoll2}; unfortunately, with the current data the error bars are large in this case (see the second-from-rightmost points in Figure 4 of \cite{kn:STARcoll}), and clear evidence of polarization is yet to be obtained. There is in any case no conflict with our claim that $\overline{\mathcal{P}}_{\Lambda'}\,+\,\overline{\mathcal{P}}_{\overline{\Lambda}'}$ can be no larger than this or indeed that (\ref{BETA}) might be valid here.

At the other extreme, one can consider the lead-lead collisions studied in the ALICE experiment at the LHC: here, in the collisions at 2.76 TeV, the energy density \cite{kn:aliceenergy} is about 2.3 times larger than in the 200 GeV collisions, but the angular momentum density is about 13.5 times larger for a given centrality; furthermore, the temperature is considerably higher, roughly 300 MeV. The ALICE investigation of peripheral collisions \cite{kn:bed} considered centrality in two ranges: $15\%$ to $50\%$, and also $5\%$ to $15\%$. As before, in the first case we can take $20\%$ to be representative, and then we obtain from (\ref{ALPHA}) an extremely severe bound:
\begin{equation}\label{S}
\left[\overline{\mathcal{P}}_{\Lambda'}\,+\,\overline{\mathcal{P}}_{\overline{\Lambda}'}\right]\left(\sqrt{s_{\m{NN}}} = 2.76\, \m{TeV},\,\mathcal{C} = 20\%\right) \;\leq \;\approx\, 0.046\%.
\end{equation}
The other range is interesting, since the data go down to a very low centrality. Here the angular momentum is enormous, but it does not vary monotonically with impact parameter, so this case merits separate investigation. The much larger overlap volume when the impact parameter is small (around 3.5 fm for $5\%$ centrality) makes itself felt here, and we find in this case a slightly \emph{less} stringent bound despite the higher angular momentum:
\begin{equation}\label{T}
\left[\overline{\mathcal{P}}_{\Lambda'}\,+\,\overline{\mathcal{P}}_{\overline{\Lambda}'}\right]\left(\sqrt{s_{\m{NN}}} = 2.76\, \m{TeV},\,\mathcal{C} = 5\%\right) \;\leq \;\approx\, 0.055\%.
\end{equation}
This interesting relaxation of the bound at low centralities is characteristic of the holographic model, and we will discuss it in more detail elsewhere. For the present we merely note that this is still an extremely low value.

Even with the substantial (theoretical and observational) uncertainties here, it is clear that, in all cases, the vorticity bound is (at present) completely inconsistent with any observation of hyperon polarization in these experiments (and of course this prediction is even more firm for the collisions at 5.02 TeV \cite{kn:ALICEoverview})\footnote{Hyperon polarization may, however, be observable at very high impact energies in future, perhaps in runs 3 or 4 of the LHC \cite{kn:future}.}. This is entirely consistent with the reported data, in which no evidence of $\Lambda/\overline{\Lambda}$ polarization was detected \cite{kn:bed}.

In summary, the vorticity bound asserts that global polarization of $\Lambda$ and $\overline{\Lambda}$ hyperons should certainly not be observable in current data at impact energies much above 200 GeV. It is consistent with a tiny total polarization at 200 GeV ---$\,$ now observed, at almost exactly the maximum value permitted by the bound. If the uncertainties can be very considerably reduced, and if (\ref{BETA}) continues to hold, we expect it to be observable in collisions at 62.4 GeV, at a total percentage about double the observed value at 200 GeV.

Let us turn, then, to much \emph{lower} impact energies.

\subsubsection*{{\textsf{5.2 Collisions at 39, 27, and 19.6 GeV }}}
The STAR collaboration took data at 39, 27, and 19.6 GeV impact energies. We interrupt our investigation at 19.6 GeV because, while data were also taken at still lower impact energies (to be discussed below), it is not completely clear that the QGP is actually formed in those cases; this is discussed in detail in \cite{kn:sahoo}. We will not take a stand on this issue, but we find it clearest to focus first on the cases which are not in doubt.

In the case of collisions at 39 GeV, with $20\%$ centrality, we find that the angular momentum density $\alpha$ has dropped to around $147.8/\m{fm}^3$, the energy density $\varepsilon$ to $7.25/\m{fm}^4$, the temperature to 178 MeV, and so the vorticity bound (\ref{ALPHA}) gives us
\begin{equation}\label{U}
\left[\overline{\mathcal{P}}_{\Lambda'}\,+\,\overline{\mathcal{P}}_{\overline{\Lambda}'}\right]\left(\sqrt{s_{\m{NN}}} = 39\, \m{GeV},\,\mathcal{C} = 20\%\right) \;\leq \;\approx\, 1.51\%;
\end{equation}
the corresponding collisions at 27 GeV have $\alpha \approx 102.3/\m{fm}^3$, $T \approx 172$ MeV, and $\varepsilon \approx 5.89/\m{fm}^4$, and so we have
\begin{equation}\label{UU}
\left[\overline{\mathcal{P}}_{\Lambda'}\,+\,\overline{\mathcal{P}}_{\overline{\Lambda}'}\right]\left(\sqrt{s_{\m{NN}}} = 27\, \m{GeV},\,\mathcal{C} = 20\%\right) \;\leq \;\approx\, 1.83\%.
\end{equation}

Finally, for collisions at 19.6 GeV we have a still lower angular momentum density of around $74.2/\m{fm}^3$, $T \approx 171$ MeV, and the energy density is about $5.6/\m{fm}^4$, leading to
\begin{equation}\label{V}
\left[\overline{\mathcal{P}}_{\Lambda'}\,+\,\overline{\mathcal{P}}_{\overline{\Lambda}'}\right]\left(\sqrt{s_{\m{NN}}} = 19.6\, \m{GeV},\,\mathcal{C} = 20\%\right) \;\leq \;\approx\, 2.42\%.
\end{equation}

The agreement with Figure 4 of \cite{kn:STARcoll} (sixth pair from left for 39 GeV, fifth from left for 27 GeV, fourth from left for 19.6 GeV) (of course one has to add the two values shown there at each impact energy) is better than one was entitled to expect in a holographic model (that is, agreement to within a factor of at best 2). The rate at which the total polarization declines with increasing impact energy is reproduced particularly well.

In short: at these impact energies, the vorticity bound relaxes quite dramatically, to the point where global polarization of $\Lambda$ and $\overline{\Lambda}$ hyperons should be clearly observable; and so it has proved: these are the impact energies for which the evidence for hyperon polarization arising from QGP vorticity was most clear-cut in \cite{kn:STARcoll}.

Finally, we consider the collisions with the lowest impact energies.

\subsubsection*{{\textsf{5.3 Collisions at 14.5, 11.5, and 7.7 GeV}}}
The reported data \cite{kn:STARcoll} on the $\Lambda$ and $\overline{\Lambda}$ hyperon polarizations present a less clear picture than in the case just considered. In particular, the $\overline{\Lambda}$ polarization results appear to be significantly larger than those for $\Lambda$ hyperons, and this suggests that some additional effect may be at work here, making the interpretation of these results somewhat dubious: see \cite{kn:kolo} and particularly \cite{kn:csernkap}. In addition, at these impact energies (particularly for the 7.7 GeV case), it is open to doubt whether a QGP actually forms. If this is not the case, of course, then a gauge-gravity approach \emph{cannot be used}.

With these warnings noted, the predictions of the holographic model are as follows.

At 14.5 GeV, $\alpha \approx 54.96/\m{fm}^3$, $T \approx 168$ MeV, $\varepsilon \approx 4.56/\m{fm}^4$, and then
\begin{equation}\label{W}
\left[\overline{\mathcal{P}}_{\Lambda'}\,+\,\overline{\mathcal{P}}_{\overline{\Lambda}'}\right]\left(\sqrt{s_{\m{NN}}} = 14.5\, \m{GeV},\,\mathcal{C} = 20\%\right) \;\leq \;\approx\, 2.73\%;
\end{equation}
collisions at 11.5 GeV have $\alpha \approx 43.59/\m{fm}^3$, $T \approx 164$ MeV, and $\varepsilon \approx 3.97/\m{fm}^4$, leading to
\begin{equation}\label{X}
\left[\overline{\mathcal{P}}_{\Lambda'}\,+\,\overline{\mathcal{P}}_{\overline{\Lambda}'}\right]\left(\sqrt{s_{\m{NN}}} = 11.5\, \m{GeV},\,\mathcal{C} = 20\%\right) \;\leq \;\approx\, 3.04\%;
\end{equation}
and finally the 7.7 GeV collisions have $\alpha \approx 29.18/\m{fm}^3$, $T \approx 160$ MeV, and $\varepsilon \approx 3.00/\m{fm}^4$, giving
\begin{equation}\label{Y}
\left[\overline{\mathcal{P}}_{\Lambda'}\,+\,\overline{\mathcal{P}}_{\overline{\Lambda}'}\right]\left(\sqrt{s_{\m{NN}}} = 7.7\, \m{GeV},\,\mathcal{C} = 20\%\right) \;\leq \;\approx\, 3.53\%.
\end{equation}

Except at 7.7 GeV, the agreement with \cite{kn:STARcoll} continues to be fairly good. In the 7.7 GeV case, the reported polarization for $\overline{\Lambda}$ hyperons is so much larger than that for $\Lambda$ hyperons that this case should be viewed with particular caution. In any event, in view of the large error bars in these cases, we can still assert that there is at least no contradiction to the vorticity bound.

It is noteworthy that, as one proceeds to higher impact energies, the difference between the reported $\Lambda$ and $\overline{\Lambda}$ hyperon polarizations grows steadily smaller, being quite negligible \cite{kn:STARcoll2} at 200 GeV; at the same time, the agreement of the vorticity bound, and of equation (\ref{BETA}), with the data becomes steadily better. This may not be a coincidence.

\begin{figure}[!h]
\centering
\includegraphics[width=1\textwidth]{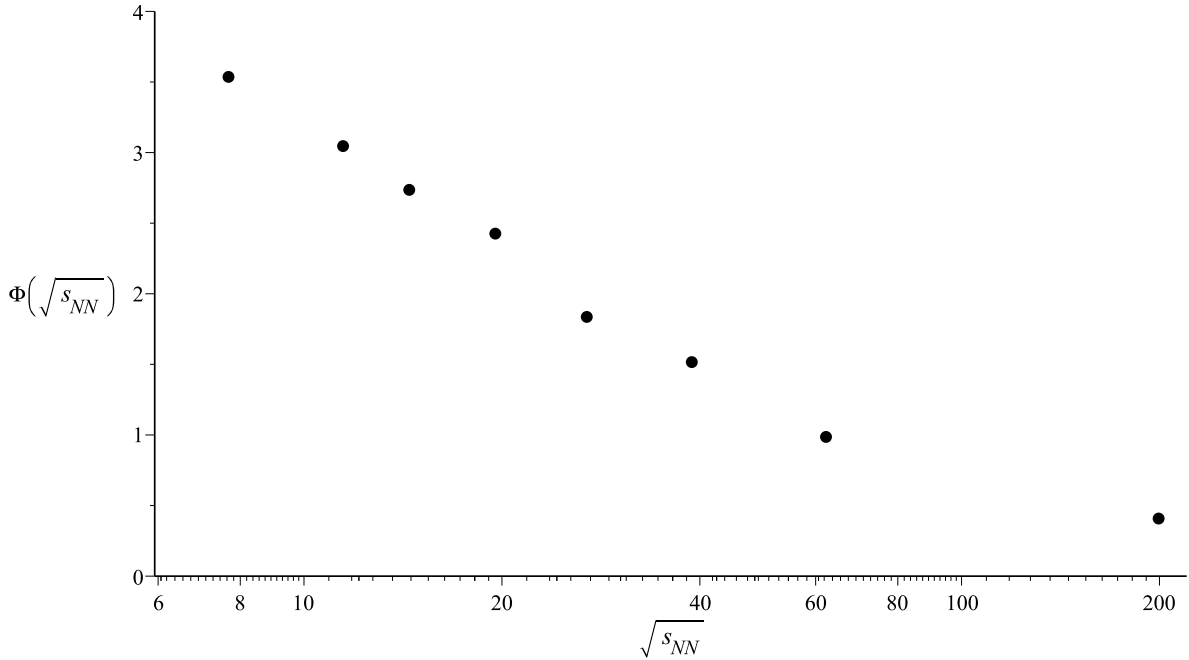}
\caption{Theoretical upper bounds on total $\Lambda$ hyperon polarization, that is, $ \left[\overline{\mathcal{P}}_{\Lambda'}\,+\,\overline{\mathcal{P}}_{\overline{\Lambda}'}\right]\left(\sqrt{s_{\m{NN}}},\,\mathcal{C} = 20\%\right) \,\leq \,  \Phi\left(\sqrt{s_{\m{NN}}},\, \mathcal{C} = 20 \%\right)$, as a percentage, for collisions at $\sqrt{s_{\m{NN}}} = 7.7,\, 11.5,\, 14.5,\, 19.6,\, 27,\, 39,\, 62.4,\, 200$ GeV and $20\%$ centrality.}
\end{figure}

Our results are summarized in Figure 1, which should be compared with Figure 4 of \cite{kn:STARcoll} and Figure 4 of \cite{kn:STARcoll2} by adding together the values corresponding to the two points at each impact energy. The figures appear to be compatible.

A more broad-brush way of making a comparison with the results of \cite{kn:STARcoll} is to compute the vorticity itself, averaged over impact energies. As mentioned above, in \cite{kn:STARcoll} this is given as $9\,\pm 1\,\times 10^{21}\,\cdot\,$s$^{-1}$, but with a large systematic uncertainty of order 2. Here we find that the $\sqrt{s_{\m{NN}}}$-averaged value of $\omega$, computed using (\ref{BETA}), is approximately $5.3\,\times 10^{21}\,\cdot\,$s$^{-1}$, somewhat low, but in reasonable agreement with the data in view of the uncertainties. (The principal uncertainty is, once again, primarily associated with the difficulty \cite{kn:bus} of determining the temperatures; the temperature estimates used in \cite{kn:STARcoll} differ somewhat from those used here.)

Our claim, then, is that the relation (\ref{BETA}), inspired by the simplest possible holographic model of this system, approximately captures the actual relation between the vorticities and the angular momentum densities of the plasmas generated by peripheral collisions, at least for impact energies which are not very low (meaning below 11.5 GeV).

We should also be cautious with regard to centralities. We have seen that both (\ref{ALPHA}) and (\ref{BETA}) are valid for collisions at $\sqrt{s_{\m{NN}}} = 11.5$ GeV and centrality $20\%$, with $\alpha \approx 44$ fm$^{-3}$. We should therefore not assume that the bound is attained at any angular momentum density below around $40$ fm$^{-3}$. This translates to an impact parameter no lower than 2.5 fm (or centrality about 2.5$\%$). On the other hand, all of our discussions have concerned collisions which are not very peripheral, with centrality not much greater than 20$\%$, corresponding to an impact parameter no greater than about 7 fm. This, then, is the domain in which we claim that (\ref{ALPHA}) and (\ref{BETA}) are valid.

\section* {\large{\textsf{6. Conclusion}}}
We have studied the AdS$_5$-Kerr spacetime from a holographic point of view. Such a black hole, with an angular momentum to mass ratio $\mathcal{A}$, corresponds to matter at conformal infinity with an angular momentum density to energy density ratio also equal to $\mathcal{A}$, and with an angular velocity which can at least be bounded above. We have conjectured that a more complete analysis, were it possible, would turn this bound into an equation, and we have argued that the data reported by the STAR collaboration is consistent with this conjecture; so are the corresponding results from ALICE at the LHC, in the sense that the non-observation of $\Lambda$/$\overline{\Lambda}$ hyperon polarization there is consistent with the small values predicted by equation (\ref{BETA}).

The applicability of holographic techniques to this problem is fundamentally limited: the no-hair theorems ensure that we have very few parameters at our disposal in the bulk. The ``universality'' of black hole physics is often cited \cite{kn:nat} as a virtue of the holographic approach, but in this case it severely restricts the number of properties of the ``peripheral plasma'' we can hope to represent\footnote{The only parameter we have not used is the angular momentum corresponding to rotation of the bulk black hole around a second axis; that is, one could use the most general metric given in \cite{kn:hawk}, the metric $g\left(\m{AdSK}_5^{(a,b)}\right)$ in our notation, where $b$ represents a second, independent angular momentum parameter. More speculatively, one could try to use five-dimensional rotating objects with non-spherical horizon topologies, if these can be found explicitly in the asymptotically AdS context \cite{kn:reall}.}.

An optimistic assessment of these results would assert that, within its domain of applicability, the holographic model works unexpectedly well. The agreement of Figure 1 with Figure 4 of \cite{kn:STARcoll} and Figure 4 of \cite{kn:STARcoll2}, apart from one possible outlier, is surprising. The fact that the model predicts, correctly, that hyperon polarization associated with QGP vorticity should be readily observable at impact energies up to around 39 GeV, observable only with difficulty at impact energy 200 GeV, and not at all (in current experiments) at higher energies, is very suggestive. A pessimistic assessment would assert that the predictions of the holographic model are at least not in blatant conflict with the data.

Even if one is sceptical regarding the holographic model, the results do make it reasonable to conjecture that vorticity in the QGP is subject to some kind of general constraint, and that its mathematical form is similar to that of (\ref{ALPHA}) or  (\ref{BETA}). Our simple model produces a very specific value for the constant $\varkappa\,$ occurring in those relations; perhaps this can be improved or given a firmer basis by more sophisticated considerations. At least we have a concrete basis for further investigations by other methods.

We have seen that holography focuses our attention\footnote{It might be said \cite{kn:karch} that focusing attention on the ``right'' variables is one of the principal services that holography can render.} on a specific parameter, the ratio $\varepsilon/\alpha$. This quantity depends in a complicated but definite manner on the centrality of a peripheral collision, and the dependence is particularly marked for centralities much smaller than those considered here (or in \cite{kn:STARcoll,kn:STARcoll2}). Our considerations therefore allow predictions to be made regarding what one must expect to find if data can be taken at small centralities. This will be discussed elsewhere.

\addtocounter{section}{1}
\section*{\large{\textsf{Acknowledgements}}}
The author thanks Dr Soon Wanmei for valuable discussions.

\end{document}